\documentclass[twocolumn,a4paper,8pt]{article}
\usepackage[english]{babel}
\usepackage{url}
\usepackage{mathrsfs}
\usepackage{multicol}
\usepackage{wrapfig}
\usepackage{graphicx}
\usepackage{flushend}
\usepackage{amsmath}
\usepackage[utf8]{inputenc} 
\usepackage{vmargin}
\usepackage{amsfonts}
\usepackage{amssymb}
\usepackage{graphicx,wrapfig,lipsum}
\usepackage{caption}
\usepackage{subcaption}
\usepackage{xcolor}
\usepackage{multirow}
\usepackage{vmargin}
\usepackage{enumitem}
\setmargins{2cm}   
{2cm}       
{17cm}      
{20cm}     
{10pt}        
{0.5cm}         
{3pt}         
{2cm}         

\addto\captionsenglish{}

\def\bra#1{\mathinner{\langle{#1}|}}
\def\ket#1{\mathinner{|{#1}\rangle}}

\def\bravert{\egroup\,\vrule\,\bgroup}

\newcommand{\vac}{\ket{\text{vac}}}
\newcommand{\vacbra}{\bra{\text{vac}}}

\begin{document}

\title{An integrated photonic circuit for color qubit preparation by third-order nonlinear interactions}
\author{A. L. Aguayo-Alvarado$^{1*}$, F. Dom\'inguez-Serna$^{2*}$, W. De La Cruz$^3$, K. Garay-Palmett$^{1*}$ 
\vspace*{0.3cm}
\\ $^1$Departamento de \'Optica - Centro de Investigaci\'on Cient\'ifica y de Educaci\'on Superior de \\ Ensenada, B.C., 22860, M\'exico 
\\ $^2$ C\'atedras Conacyt - Centro de Investigaci\'on Cient\'ifica y de Educaci\'on Superior de \\ Ensenada, B.C., 22860, M\'exico
\\ $^3$ Centro de Nanociencias y Nanotecnolog\'ia, Universidad Nacional Aut\'onoma de M\'exico, \\ Km. 107 Carretera Tijuana-Ensenada, 22860, B.C., M\'exico 
\\ *e-mails: aaguayo@cicese.edu.mx, fadomin@cicese.edu.mx, kgaray@cicese.edu.mx
}

\date{ }

\twocolumn[
  \begin{@twocolumnfalse}
    \maketitle
    \begin{abstract}
This work presents a feasible design of an integrated photonic circuit performing as a device for single-qubit preparation and rotations through the third-order nonlinear process of difference frequency generation (DFG) and defined in the temporal mode basis. The first stage of our circuit includes the generation of heralded single photons by spontaneous four-wave mixing in a micro-ring cavity engineered for delivering a single-photon state in a unique temporal mode. The second stage comprises the implementation of DFG in a spiral waveguide with controlled dispersion properties for reaching color qubit preparation fidelity close to unity. We present the generalized rotation operator related to the DFG process, a methodology for the device design, and qubit preparation fidelity results as a function of user-accessible parameters. 
\normalsize
\vspace*{0.5cm}
 \end{abstract}
  \end{@twocolumnfalse}
]

\section*{INTRODUCTION}

Quantum information processing (QIP) involves manipulating and processing information coded on qubits. Several two-level systems have been proposed and implemented for qubit preparation, such as trapped ions \cite{Debnath2016}, quantum dots \cite{Loss1998,Uppu2020}, and superconducting materials \cite{Song2017}. However, single photons have been one of the most studied systems due to their more straightforward implementation, and relative lack of decoherence \cite{Ladd2010}. Broadly, these single-photon states are generated from parametric nonlinear processes, such as spontaneous parametric downconversion \cite{Mosley2008} and spontaneous four-wave mixing \cite{Soller2011}. Practical implementations of QIP tasks, like quantum computing, demand scalability of the quantum systems \cite{DiVicenzo2000}, for which integrated micro and nanodevices become an attractive alternative \cite{Qiang2018,Zhukovsky2012, Signorini2020}. 

In generating single-photon-based qubits, the polarization basis has been the most exploited \cite{Wang2019,Crespi2011,Pewkhom2020,Laplane2016}; though, its two-dimensional Hilbert space does not allow harnessing the real potential of QIP. In recent years, different bases for qubit preparation have emerged as alternatives to the polarization basis. For example, the transverse spatial degree of freedom, through the orbital angular momentum (OAM) basis \cite{Nicolas2014}, and the temporal/spectral degree of freedom in terms of the temporal mode (TM) basis \cite{Raymer2020}. Both schemes have attracted significant attention to encoding information since they reside in high-dimensional Hilbert spaces. 

The temporal mode basis has advantages over the OAM basis since it can be used from pulsed optical fields in guided media and is less sensitive to disturbances \cite{Raymer2020,Brecht2015}, establishing an excellent option for applications in QIP \cite{Ashby2020}. Recent work on quantum pulse gate \cite{Eckstein2011,Brecht2011, Allgaier2020} has opened up a route for TM quantum logic implementation based on nonlinear interactions that involve frequency conversion of optical fields as an alternative to the extensively implemented quantum computing protocols by linear optics \cite{Knill2001}. Temporal mode based qubits find applications, for instance, in quantum network nodes, where the frequency of flying qubits (photon states) can be converted (through a unitary transformation resulting from a nonlinear process) to a different frequency at which the stationary qubits (usually a material system) are sensitive \cite{Maring2018}, but without altering the information coded in their quantum state \cite{McKinstrie2005,McGuinness2010}. The former kind of qubits is known as color qubits \cite{Clemmen2016,Dominguez2021}, in the sense that the quantum state can be in one frequency or another, i.e., there is a two-level system that allows for the interaction of two frequencies $\omega_s$ and $\omega_r$, whose linear combination forms a qubit of the form $|\Psi\rangle = c_r|\omega_r\rangle + c_s|\omega_s\rangle$, where $|\omega_j\rangle$ is a single-photon state of frequency $\omega_j$ and $c_j$ its corresponding complex coefficient, with $j =  {s,r}$.

In this work, we propose a feasible integrated photonic circuit, capable of color qubit preparation (in the TM basis) and rotations by the third-order nonlinear process of difference frequency generation (DFG) \cite{Fewings2000}. Through this process, qubit rotations are restricted to axes lying in the $xy$ plane of the Bloch sphere.  The present work constitutes a step forward in the study that we recently published \cite{Dominguez2021}. The proposed device (see Figure \ref{fig:Scheme}) constitutes an example with high potential to be implemented, in which the demonstration of color qubit preparation by DFG with fidelities close to unity is achievable. The proposal comprises the on-chip integration of a heralded single-photon source by SFWM in a micro-ring cavity and a DFG source implemented in a spiral waveguide. Both sources have been engineered to increase the spectral overlap between the fundamental temporal modes of the input state (based on SFWM) and the DFG quantum gate, which lead to the desired qubit preparation fidelity. There are alternative mechanisms for controlling the spectral properties of the involved processes, such as those based on the group velocity matching technique \cite{Grice2001,Garay2007}, through which similar results could be obtained and could also be implemented in integrated photonic circuits. In this sense, the presented theory and design methodology are general and can be applied in several configurations. Note that the TM basis for color qubit preparation demands unitary conversion efficiency and mode selectivity. However, in practical implementations reaching unitary efficiencies involve time-ordering corrections (ignored in the present treatment) \cite{Reddy2014,Reddy2015} that degrade the selectivity. This limitation can be overcome by implementing a cascading scheme based on an optical Ramsey interferometer \cite{Quesada2016, Reddy2018}, a scenario we shall consider as future work.

\section*{RESULTS}

\begin{figure*}[t!]
  \begin{center}
  		\includegraphics[width=0.9\linewidth]{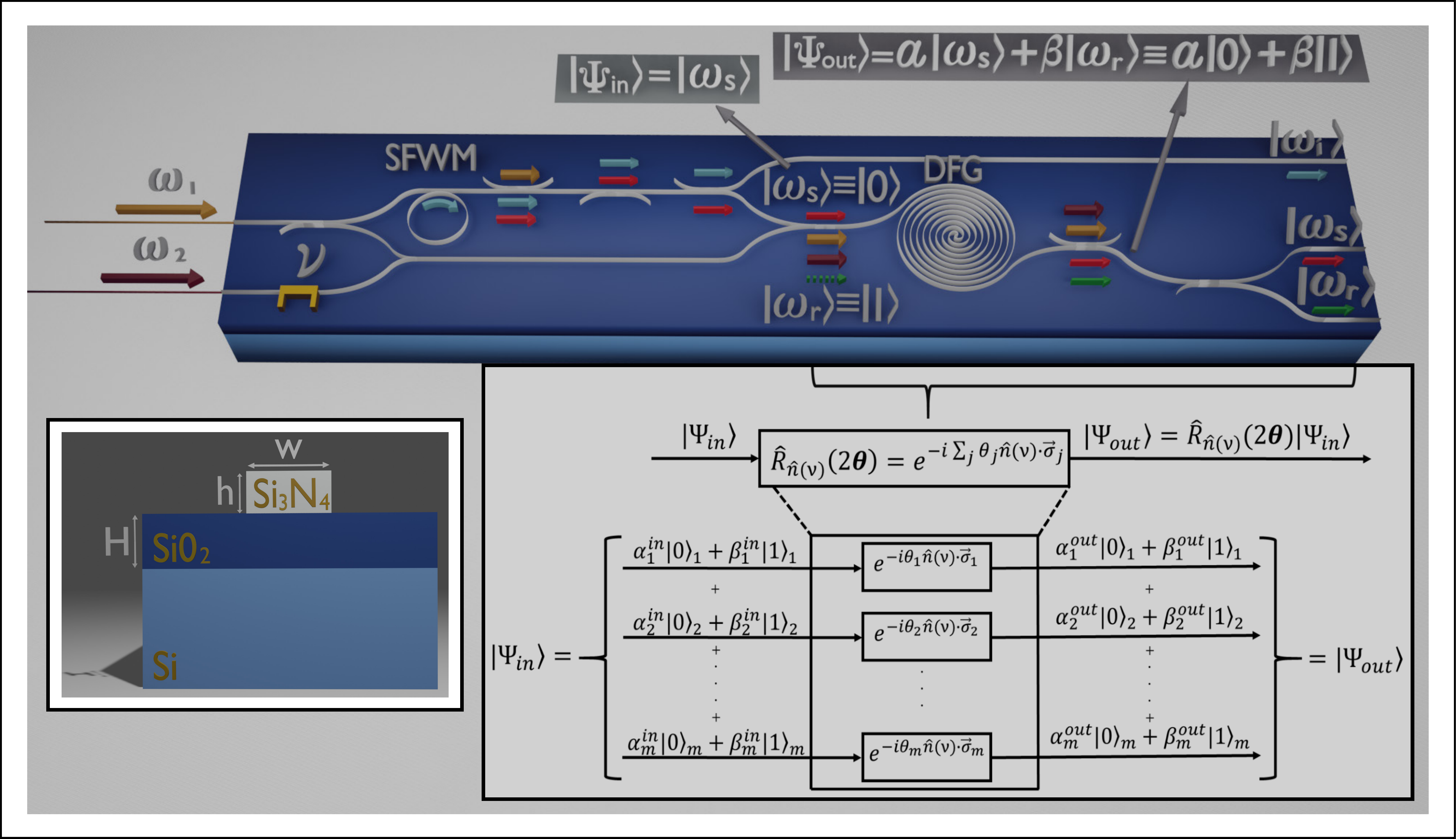}
  	\caption{\footnotesize{Design of the proposed integrated photonic circuit for color qubit preparation and rotations. All components in the device are based on silicon nitride (Si$_3$N$_4$) waveguides above a substrate of silicon dioxide (SiO$_2$) on silicon (Si). The top scheme describes the circuit operation, comprising the stage for generating a heralded single-photon state, which acts as input to the quantum gate waveguide (the second stage), where the DFG process occurs. The bottom-right scheme depicts the DFG process that leads to a generalized rotation operator for single-qubits defined in the temporal mode basis. Note that $\ket{0}= \ket{ \phi_j}$  and  $\ket{1} = \ket{ \psi_j}$ are written on a computational basis in pairs of temporal modes. The bottom-left image is a representation of the waveguide's geometry.}}
  	\label{fig:Scheme}
  	\end{center}
  \end{figure*}

This paper focuses on the third-order difference frequency generation process regarded as a unitary evolution for single-qubit preparation and rotations, which we propose implementing in the integrated photonic circuit of Figure \ref{fig:Scheme}, designed under realistic conditions. We prove that the DFG process leads to a generalized rotation operator, around an axis with unit vector $\hat{\bold{n}}(\nu)$, in the form $\textbf{R}_{\hat{\bold{n}}(\nu)}=\mbox{exp}\left(-i\sum_{j}\theta_j\hat{\bold{n}}(\nu)\cdot \hat{\boldsymbol{\sigma}}_j \right)=\prod_jR^j_{\hat{\bold{n}}(\nu)}(2\theta_j)$, where
\begin{equation}
	R^j_{\hat{\bold{n}}(\nu)}(2\theta_j)=\mathbb{\textbf{I}}-\mathbb{I}_j+\mbox{cos}\theta_j\mathbb{I}_j-i\mbox{sin}\theta_j\left(\hat{\bold{n}}(\nu)\cdot \hat{\boldsymbol{\sigma}}_j \right),
	\label{eq:rotOper}
\end{equation}
\noindent with $\hat{\boldsymbol{\sigma}}_j$ a vector of Pauli matrices, $\mathbb{\textbf{I}}$ the identity matrix, $\mathbb{I}_j=\ket{\phi}\!_{j\,j}\!\bra{\phi} + \ket{\psi}\!_{j\,j}\!\bra{\psi}$, and $\hat{\bold{n}}(\nu)=\langle\mbox{cos}\,\nu,\mbox{sin}\,\nu,0\rangle$. Each matrix $\hat{\sigma}^j_{x,y,z}$ couples the subspace $j$, $\{\ket{\phi}_j, \ket{\psi}_j\}$, such that each $R^j_{\hat{\bold{n}}(\nu)}(2\theta_j)$ is valid for a pair of modes ($\ket{\phi}_j, \ket{\psi}_j$). For instance, $\hat{\sigma}_x$ in subspace $j$ is written as $\hat{\sigma}_x^j = \vert\psi\rangle_j {}_j \langle\phi \vert+ \vert\phi\rangle_j {}_j\langle\psi\vert$. Note $\textbf{R}_{\hat{\bold{n}}(\nu)}$ describes a rotation around an axis lying in the xy plane. The axis is defined by $\nu$, while the rotation angle $\theta_j$ varies for each $j$, which means that once the rotation axis is defined, all involved pairs ($\ket{\phi}_j, \ket{\psi}_j$) will rotate to a different angle. Therefore, after evolution, the output state will comprise in a linear combination several unwanted pairs, unless a single one interacts in a controlled manner \cite{Dominguez2021}, a challenge we endeavored to overcome. Note that our treatment relies on the description of pulsed single-photon states in the temporal mode basis, such that $\ket{\phi}_j$ and $\ket{\psi}_j$ become to represent single-photon states in a single temporal mode. Also, note that in our proposal $\hat{\bold{n}}(\nu)$ can be externally controlled, while $\theta_j$ will depend on the device geometry and experimental parameters. 

Following the scheme in Figure \ref{fig:Scheme}, the designed photonic circuit involves the generation of a heralded single-photon state by spontaneous four-wave mixing in a micro-ring resonator. The SFWM source is engineered for delivering a single-temporal-mode photon state $\ket{\omega_s}$, which becomes one of the two states of the computational basis (say $\ket{0}$). Then, this state is launched together with two pulsed pump fields (tuned at frequencies $\omega_1$ and $\omega_2$) into a waveguide, dispersion-controlled to fulfill phasematching for DFG at the desired frequencies. DFG can be defined as a frequency-translation process leading to the annihilation of $\omega_s-$photons and subsequent emission of single-photons at $\omega_r$, or the corresponding reversed interaction. The frequency conversion probability will depend on source parameters as the power of the pump fields.  Note that in this description, a single-temporal-mode state $\ket{\omega_r}$ would correspond to the other state of the computational basis ($\ket{1}$). 

The proposed DFG source is designed to favor the interaction in a particular sub-space $j$, i.e., to mediate the coupling between the temporal-mode pair ($\ket{\phi}_j, \ket{\psi}_j$). In such a case, even the DFG process involves the emission of a linear combination of temporal-modes pairs (as seen in figure \ref{fig:Scheme}), only the one matching the temporal mode of the single-photon input state will be efficiently delivered from the waveguide end. Roughly, the DFG nonlinear interaction can be described as a unitary evolution $\hat{U}$ that produces color qubits in the form $\ket{\Psi}_{out}=\hat{U}\ket{\omega_s}=\alpha\ket{\omega_s}+\beta \ket{\omega_r}$, with $|\beta|^2$ the probability of conversion and $|\alpha|^2=1-|\beta|^2$ \cite{Dominguez2021}. However, such description corresponds to the ideal situation for which the DFG unitary operator involves a unique temporal-mode pair ($\ket{\phi}_j, \ket{\psi}_j$), and also the input state. In general, both the input single-photon state and the DFG source comprise a linear superposition of temporal modes, which lead to unwanted interactions through the evolution in the nonlinear medium.

Our proposal is based on spectral-properties control engineering to let DFG interactions serve as a platform for qubit preparation and rotations. To quantify in a realistic scenario how close is the evolved state to the ideal one, we use a fidelity measure defined as $\mathcal{F} =\,^{\text{ideal}}_{\text{out}}\!\!\bra{\Psi}\rho_{out}\ket{\Psi}_{\text{out}}^{\text{ideal}}$, where $\rho_{out}$ is the density operator denoting the evolved state.

Focusing on the ideal unitary transformation sketched in the top scheme of Figure \ref{fig:Scheme}, the input state can be already in a superposition $\alpha\ket{\omega_s}+\beta\ket{\omega_r}$, previously prepared in an identical DFG source, which could be integrated to the same photonic circuit. On the other hand, let us highlight that once a color qubit is available, rotations, given in Equation (\ref{eq:rotOper}), are achievable for particular values of the phase difference ($\nu$) between the two pump fields. Thus, our proposal has the potential to implement, in an integrated monolithic device, the generation of a single-photon state, the preparation of qubits, and their rotations on the Bloch sphere.

\subsection*{The proposed device}
We propose an integrated photonic circuit for color-qubit quantum gate implementation based on silicon nitride rectangular waveguides lying above a plane substrate of silicon dioxide on silicon, as shown in the inset of Figure \ref{fig:Scheme} \cite{Aguayo2020a}. The device integrates the stages for single-photon state generation by SFWM and the qubit preparation/rotation by DFG.

\begin{table}[t]
\centering
\caption{\bf Parameters of the SFWM and DFG sources of the proposed integrated photonic circuit}
\begin{tabular}{ccc}
\hline 
 & SFWM & DFG\\
\hline
\multicolumn{3}{c}{Central wavelengths ($\mu$m)} \\ \hline
Pump-1&$\lambda_{1}=0.822$ & $\lambda_{1}=0.822$\\
Pump-2& & $\lambda_{2}=1.554$\\
Signal-s&$\lambda_{s}=1.253$& $\lambda_{s}=1.253$\\
Signal-r& & $\lambda_{r}=0.729$\\
Idler&$\lambda_{i}=0.612$& \\ \hline
\multicolumn{3}{c}{Waveguide dimensions ($\mu$m)} \\ \hline
Core width&$w_{sfwm}=0.953$ & $w_{dfg}=1.617$\\
Core height&$h=0.700$ & $h=0.700$\\
SiO$_2$ height&$H=1.0$ & $H=1.0$\\
Length&$l_c=43$ & $L=1.0\times 10^4$\\
\hline
\multicolumn{3}{c}{Pump/filter bandwidths (THz)} \\ \hline
Pump-1&$\sigma_1=6.0\,$ & $\sigma_1=6.0\,$\\
Pump-2& & $\sigma_2=0.7\,$\\
Filter&  $\sigma_f=1.0\,$&\\  \hline
\multicolumn{3}{c}{Cavity reflectivity } \\ \hline
 & $R_i=0.86$&\\  \hline
 \multicolumn{3}{c}{Nonlinear coefficient (mW)$^{-1}$} \\ \hline
 &$\gamma_{fwm}=5.05$ &$\gamma_{dfg}=2.50$ \\
\hline
\end{tabular}
  \label{tab:parameter}
\end{table}

The single-photon state source consists of a micro-ring cavity resonant to the heralding mode (at frequency $\omega_i$). The design parameters for this source to herald a pure single-photon state (at frequency $\omega_s$) comprising just the fundamental temporal mode are summarized in Table \ref{tab:parameter}. In turn, the DFG source is based on a waveguide engineered for integration with the single-photon source and reaching qubit-preparation fidelities close to the unit. The specific design parameters of the DFG source are also compiled in Table \ref{tab:parameter}. Note that the two nonlinear interactions demand the use of waveguides with different widths ($w_{sfwm}$ and $w_{dfg}$), which waveguide tapers can interconnect with the minimum of losses by designing adiabatic transitions \cite{Fu2014}.

The crucial point of our proposal is the attainment of qubit preparation fidelities close to unity. In this sense, the integration of the SFWM based single-photon source with the DFG source, as depicted in Figure \ref{fig:Scheme}, becomes essential, which demands the simultaneous fulfillment of the phasematching constraint for the two nonlinear processes, at the same pump ($\omega_1$) and heralded single-photon ($\omega_s$) frequencies.

\subsubsection*{The heralded single-photon source}
The SFWM-based source is proposed in a micro-ring cavity and is modeled according to the treatment given in reference \cite{Garay2011}. In this configuration, a factorable two-photon state can be emitted if the cavity is resonant to at least one of the photons comprising the pair. In our case, we consider a cavity designed to be resonant to the idler photons, with wavelength $\lambda_i<\lambda_1$ ($\lambda_1$ being the the pump wavelength). The cavity-modified joint spectrum is composed of a set of cavity-allowed modes and, on the whole, exhibits spectral correlation, but the portion corresponding to each cavity-allowed mode is factorable. Then, by using a suitable spectral filter, we can choose a unique resonant mode (the central one in our case), obtaining a factorable two-photon state. The idler photons act as a trigger to herald the emission of signal photons, which are generally characterized by the density operator \cite{Nielsen2010,Dominguez2020}

\begin{equation}
	\begin{aligned}
		\rho_s &= \int d \omega_s \int d\omega_s' \int d \omega_i F(\omega_s, \omega_i) F^*(\omega_s',\omega_i) \\
		&\quad \times \hat{a}_s^\dagger (\omega_s) \vac \vacbra \hat{a}_s (\omega_s'), 
	\end{aligned}
	\label{eqn:heralded}
\end{equation} 

\noindent where $F(\omega_s, \omega_i)$ is the joint spectral amplitude (JSA) function of the cavity-based two-photon state and $\hat{a}(\omega) \left(\hat{a}^\dagger(\omega)\right)$ is the annihilation (creation) operator of photons at frequency $\omega$. For accessing the temporal mode content of the two-photon state, the JSA is written in terms of the Schmidt decomposition \cite{Law2000} 

\begin{equation}
	\begin{aligned}
             F(\omega_s,\omega_i) = \sum_{m=1}^{\infty} \sqrt{D_m} \varphi_m (\omega_s) \Omega_m (\omega_i), 
	\end{aligned}
	\label{eqn:Schmidt_SFWM}
\end{equation} 

\noindent with $\varphi_m (\omega_s)$ and $\Omega_m (\omega_i)$ depicting two sets of temporal-mode basis functions, which are pairwise correlated \cite{Eckstein2011}. The coefficients $D_m$ weighing the product of temporal-modes pairs are normalized such that $\sum_m\!\!D_m=1$. Note that an ideal factorable two-photon state involves only a temporal mode pair ($\varphi_m (\omega_s)$,\,$\Omega_m (\omega_i)$). In such a case, a detection event of trigger photons heralds the emission of a signal photon in a quantum-mechanically pure state; the purity defined as $p=\sum_m D_m^2$.

  \begin{figure}[t!]
  \begin{center}
  		\includegraphics[width=0.94\linewidth]{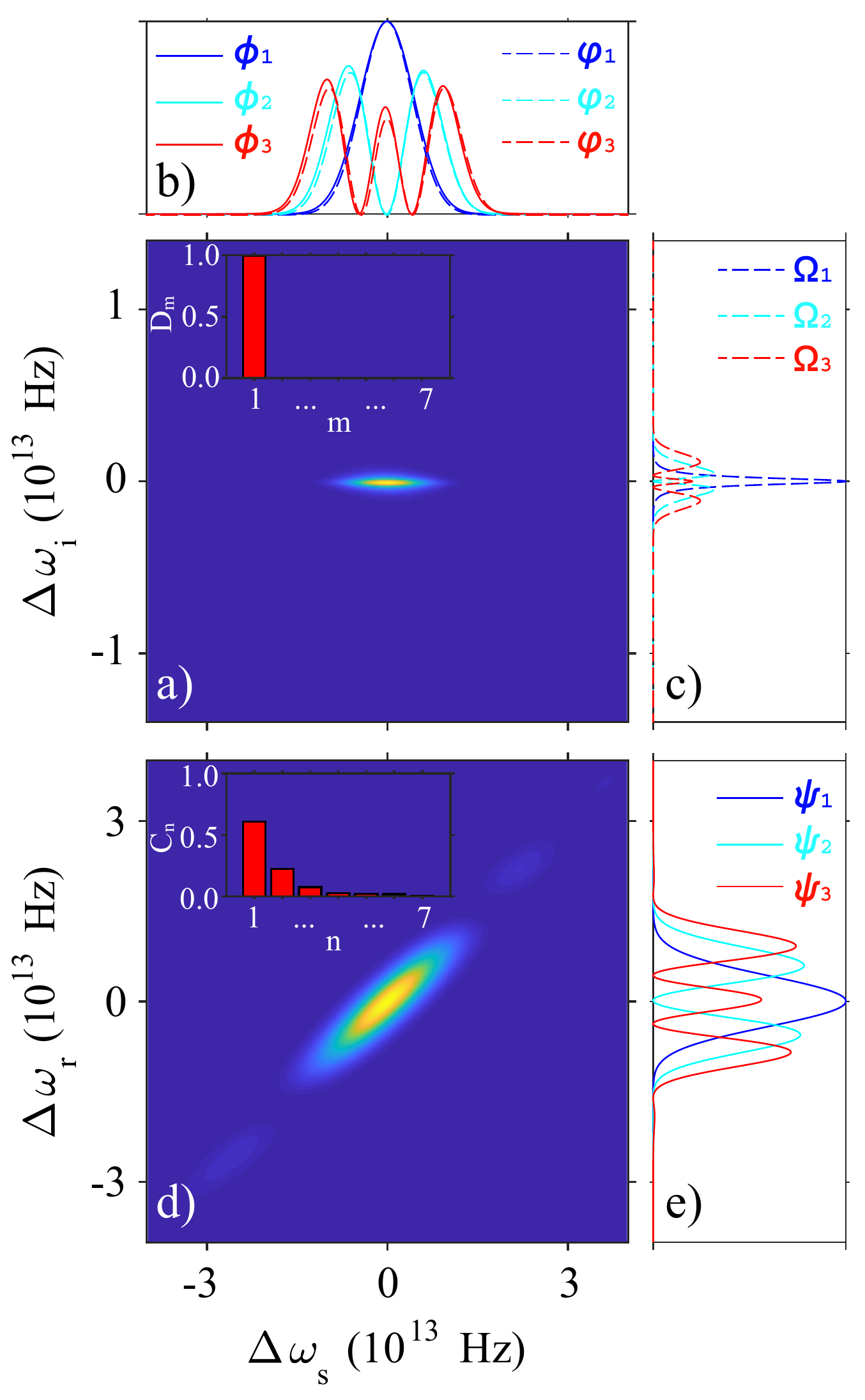}
  	\caption{\footnotesize{Spectral properties of the proposed SFWM and DFG sources. a) The normalized joint spectral intensity of the SFWM source. b) and c) The three first temporal-mode pairs of the JSA ($\varphi_m (\omega_s)$,\,$\Omega_m (\omega_i)$), plotted as dashed lines. d) The normalized DFG mapping function. b) and e) The three first temporal-mode pairs of the MF ($\phi_j (\omega_s)$,\,$\psi_j(\omega_i)$), plotted as solid lines. Note that the eigenfunctions have been normalized to the maximum value of the fundamental mode, while the joint spectral functions were normalized to their corresponding maximum value.The insets of panels a) and d) show the Schmidt coefficients of the JSA and MF, respectively.}}
  	\label{fig:spectra}
  	\end{center}
  \end{figure} 

Our proposal includes the implementation of an SFWM source in a micro-ring resonator of length $l_c$, which, after being pumped with pulses centered at $\lambda_1=2\pi c/\omega_1 = 0.822\,\mu$m is capable of delivering nearly uncorrelated photon pairs centered at wavelengths $\lambda_s=2\pi c/\omega_s = 1.253\,\mu$m and $\lambda_i=2\pi c/\omega_i = 0.612\,\mu$m. Then, detection of a visible photon (idler arm) announces an infrared photon in the signal arm, accessible for its use as input to the DFG source. The joint spectral intensity of the designed SFWM source, calculated as $|F(\omega_s, \omega_i)|^2$, is shown in Figure \ref{fig:spectra}a). Note that frequency axes are expressed as $\Delta\omega_{\mu}=\omega_{\mu}-\omega_{\mu0}$, with $\mu=s,i$ and $\omega_{\mu0}$ the central emission frequency. The joint spectrum exhibits the expected factorable feature, validated by calculating its Schmidt eigenfunctions and corresponding eigenvalues $D_m$. Figures \ref{fig:spectra}b) and \ref{fig:spectra}c) show the three first temporal-mode pairs ($\varphi_m (\omega_s)$,\,$\Omega_m (\omega_i)$), plotted as dashed lines, and the corresponding Schmidt eigenvalues are shown in the inset of Figure \ref{fig:spectra}a). As noticed, the JSA is essentially determined by the fundamental temporal-mode pair and consequently can be written as $F(\omega_s,\omega_i)\approx \varphi_1 (\omega_s) \Omega_1 (\omega_i)$. Hence, this SFWM source can lead to the preparation of a single-photon state with a spectral amplitude $\varphi_1 (\omega_s)$ and an estimated purity of $0.997$. The reached level of photon-pair source factorability was obtained by adding a spectral filter modeled as a Gaussian shape bandpass filter of bandwidth $\sigma_f=1\,$THz and centered at the central idler emission frequency of the SFWM source. This filter enables the choice of a unique cavity-allowed mode (the central one in our case) among those comprising the cavity-modified joint spectrum \cite{Garay2011} while preserving a high photon-pair emission rate per pump-power squared, $\approx0.4\,\mbox{pairs}/\mu \mbox{W}^2$. Integrated bandpass filters can be implemented from liquid crystals \cite{Xu2021}  or microresonator systems \cite{Zhang2022}.

\subsubsection*{The DFG source for qubit preparation and rotations}

The proposed DFG source is based on a $\chi^{(3)}$ spiral rectangular waveguide, as shown in Figure \ref{fig:Scheme}. In this process, a single-photon state at frequency $\omega_s$ ($\omega_r$) could be translated toward different frequency $\omega_r$($\omega_s$), through the interaction with two intense pump fields with frequencies $\omega_{1}$ and $\omega_2$ and phase difference $\nu$, constrained to energy and momentum conservation fulfillment \cite{Dominguez2021}. This $\chi^{(3)}$ nonlinear interaction is governed by an evolution operator given by
\begin{eqnarray}
	\hat{U} = \exp \left\{ \!- \zeta  \sum_{k_s} \sum_{k_r} 
	\left[ G(k_s,k_r) \hat{a}(k_s) \hat{a}^\dagger(k_r) + H.C. \right] \right\},
	\label{eqn:UConPul_1}
\end{eqnarray}

\noindent where  $\zeta$ is a global constant, $H.C.$ denotes hermitian conjugate, $k_s$($k_r$) is the propagation constant at $\omega_s$($\omega_r$), and $G(k_s,k_r) $ is the function that describes the spectral properties of the DFG process and is called onward mapping function (MF). An explicit form of this function,  given in terms of frequencies rather than wavenumbers, can be found in reference \cite{Dominguez2021}. Note that the MF squared is proportional to the probability of frequency conversion. Interestingly, the MF linearly couples the $\omega_s(k_s)$ and $\omega_r(k_r)$ modes during the interaction. The linearity of the DFG process has already been proposed as a method to obtain unitary transformations acting on qubits \cite{Brecht2015, Brecht2011}, whose computational basis is formed of two single-photon states with different colors (frequency) \cite{Nasr2021}. This feature is the one we exploit for qubit preparation and rotations in our integrated photonic device.

As for the JSA (see Equation \ref{eqn:Schmidt_SFWM}), we write the MF in terms of the Schmidt decomposition  $G(\omega_s,\omega_r) = \sum_{j=1}^{\infty} \sqrt{C_j} \phi_j (\omega_s) \psi_j (\omega_r)$, which leads us to rewrite the evolution operator of Equation (\ref{eqn:UConPul_1}) as
\begin{eqnarray}
	\hat{U} = \mbox{exp} \Big\{ -i \sum_j  \Big[ \theta_j\hat{A}_j \hat{B}^\dagger_j+ \theta_j^*\hat{A}^\dagger_j \hat{B}_j \Big] \Big\},
	\label{eqn:UConPul_2}
\end{eqnarray}

\noindent  where time-ordering effects and other possible loss mechanisms were ignored; $\phi_j (\omega_s)$[$\psi_j (\omega_r)$] depicts the Schmidt eigenfunctions, and $C_j$ is the Schmidt eigenvalue. $\hat{A}_j$  and $\hat{B}_j$ represent temporal mode annihilation operators given by

\begin{eqnarray}
\hat{A}_j = \int d\omega_s \phi_j(\omega_s) \hat{a}(\omega_s),
\label{eqn:AOper_1}
\end{eqnarray}

\begin{eqnarray}
\hat{B}_j = \int d\omega_r  \psi_j(\omega_r) \hat{a}(\omega_r).
\label{eqn:BOper_1}
\end{eqnarray}


The parameter $\theta_j$ in Equation (\ref{eqn:UConPul_2}) couples the different pairs of temporal-mode eigenstates $\ket{\phi (\omega_s)}_j\equiv\ket{\phi}_j$ and $\ket{\psi(\omega_r)}_j\equiv\ket{\psi}_j$, and is given by

\begin{eqnarray}
\theta_j = \epsilon \sqrt{C_j} L \gamma_{dfg} \sqrt{\frac{P_{av1}P_{av2}}{\sigma_1 \sigma_2}} e^{i\nu}, 
\label{eqn:thetanPulCont_2}
\end{eqnarray}

\noindent where $\epsilon$ is a global constant, $L$ is the waveguide length, $\gamma_{dfg}$ is the nonlinear coefficient, and $P_{av\mu}$ and $\sigma_{\mu}$ are the average power and bandwidth of the pump field $\mu=1,2$, respectively. Beyond experimentally controllable variables, efficient coupling between a particular temporal mode pair is determined by the Schmidt coefficient $C_j$. Equations (\ref{eqn:UConPul_2}) to (\ref{eqn:thetanPulCont_2}) reveal that the DFG process can enable the preparation of color qubits as linear combinations of the temporal-mode computational basis elements $\ket{\phi}_j$ and $\ket{\psi}_j$, in the form $\ket{\Psi}= \alpha_j \ket{\phi}_j + \beta_j \ket{\psi}_j$, see Figure \ref{fig:Scheme}.

We conducted a parametrical numerical study that enabled us to identify the geometrical and pump characteristics for the DFG source, which are summarized in Table \ref{tab:parameter}. The obtained spectral mapping function is shown in Figure \ref{fig:spectra}d), where frequency axes are also expressed in terms of detuning from the central frequencies.  Figures \ref{fig:spectra}b) and e) show the three first temporal-mode pairs ($\phi_j (\omega_s)$,\,$\psi_j (\omega_r)$), plotted as solid lines, and the first seven Schmidt eigenvalues are shown in the inset of panel \ref{fig:spectra}d). In this case, the MF exhibits marked spectral correlation as several Schmidt pairs are needed to reconstruct the entire joint spectral function. The calculated Schmidt number characterizing this source is $(\sum_jC_j^2)^{-1}=2.33$. On the other hand, note in panel b) that functions $\varphi_1(\omega_s)$ and $\phi_1(\omega_s)$ spectrally overlap almost perfectly, which was the real challenge in the quest for the design of our integrated photonic circuit. 

Let us now analyze the evolution of an input state through propagation along the DFG waveguide (quantum gate medium). Firstly, we assume a pure single-photon state as input, which can be expanded in the DFG temporal-mode basis $\{ \phi_j \}$ as $\ket{\Psi}_{in} = \sum_j \mathcal{O}_j \hat{A}^\dagger_j \vac$, where 

\begin{equation}
	\begin{aligned}
	\mathcal{O}_j=\int d\omega\phi_j^*(\omega)h(\omega),
	\end{aligned}
	\label{eq:overlap}
\end{equation}
\noindent with $h(\omega)$ the spectral amplitude of the input state, normalized as $\int\! d\omega|h(\omega)|^2=1$, that can comprise several temporal modes. In the particular example of Figure \ref{fig:Scheme}, the input state corresponds to the heralded single-photon at $\omega_s$ from the SFWM source, engineered for delivering a single-photon state in a single temporal mode, in which a case $h(\omega)$ would correspond to the first and prevailing eigenfunction in Equation (\ref{eqn:Schmidt_SFWM}), i.e., $\varphi_1(\omega_s)$.

Through the nonlinear interaction with the two pulsed pumps, the input state $\ket{\Psi}_{in}$ will evolve to the output state \cite{Dominguez2021}

\begin{equation}
	\begin{aligned}
		\ket{\Psi}_{out} &= \hat{U}\ket{\Psi}_{in} =\mathcal{N} \Big( \ket{\Psi}^{\text{ideal}}_{\text{out}} + \sum_{j\neq 1} x_j  \hat{A}_j^\dagger \vac \\ 
		& \quad+ \sum_{j\neq 1} y_j \hat{B}_j^\dagger \vac  \Big),
	\end{aligned}
	\label{eq:RealOutput}
\end{equation}
\noindent where $\mathcal{N}$ is a normalization constant and $\ket{\Psi}^{\text{ideal}}_{\text{out}} = \left( \cos \theta \,\hat{A}^\dagger -ie^{i\nu} \sin \theta\, \hat{B}^\dagger \right) \vac$, our interest's color qubit, is the evolved state resulting from a perfect spectral overlap between a temporal-mode of the input state with one of the computational basis states $\ket{\phi}_j$ or $\ket{\psi}_j$, being the fundamental mode in our particular instance, see Figure \ref{fig:spectra}b). On the other hand, the terms $x_j= \mathcal{O}_j \cos \theta_j $ and $y_j= -i\mathcal{O}_je^{i\nu}  \sin \theta_j $ arise from likely overlapping among the input state and DFG temporal-mode pairs beyond the fundamental pair. Accordingly, a linear combination of those evolved terms appears in the output state as a spurious contribution to the qubit. 

Since our proposed input state is in a unique temporal mode and the DFG source has been spectrally engineered to guarantee an almost perfect overlap between its first temporal mode  $\ket{\phi(\omega_s)}_1$ and that of $\ket{\Psi}_{in}$, the terms proportional to $x_j$ and $y_j$ in Equation (\ref{eq:RealOutput}) are negligible, enabling to our interested qubit be quite distinguishable from the spurious background. The fidelity obtained under our source parameters is $\mathcal{F}=0.99$, a highly acceptable value for practical applications. Notice that despite the mapping function comprises several eigenfunctions, the fundamental pair makes the highest contribution, as can be seen in the inset of Figure \ref{fig:spectra}d). Our results prove that it is not required to factorize the mapping function nor use external control to favor the coupling of a unique temporal mode pair ($\ket{\phi}_j$, $\ket{\psi}_j$) and lead to an almost ideal color qubit preparation. 

Let us remark that in a realistic situation, the heralded single-photon, acting as input to the quantum gate medium, will be in the mixed state $\rho_s$ given by Equation \eqref{eqn:heralded}, so that the density matrix describing the quantum gate can be calculated from Equation (\ref{eqn:UConPul_2}) as $\rho_{out} = \hat{U} \rho_{s} \hat{U}^\dagger$, obtaining

\begin{align}
	\rho_{out} &= \hat{I} \cos^2\theta |\omega_s\rangle \langle \omega_{s'}| + ie^{-i\nu}\hat{I}\cos\theta \sin\theta|\omega_s\rangle \langle \omega_r| \nonumber \\
	&- ie^{i\nu}\hat{I} \sin\theta \cos\theta |\omega_r\rangle \langle \omega_{s'}| +\hat{I}\sin^2\theta |\omega_r\rangle \langle \omega_r|,
	\label{eqn:rhos_out}
\end{align}

\noindent where $\hat{I}$ is the double integral operator. Note this expression is less intuitive than the state in Equation \eqref{eq:RealOutput}. 

  \begin{figure}[t!]
  \begin{center}
  		\includegraphics[width=0.8\linewidth]{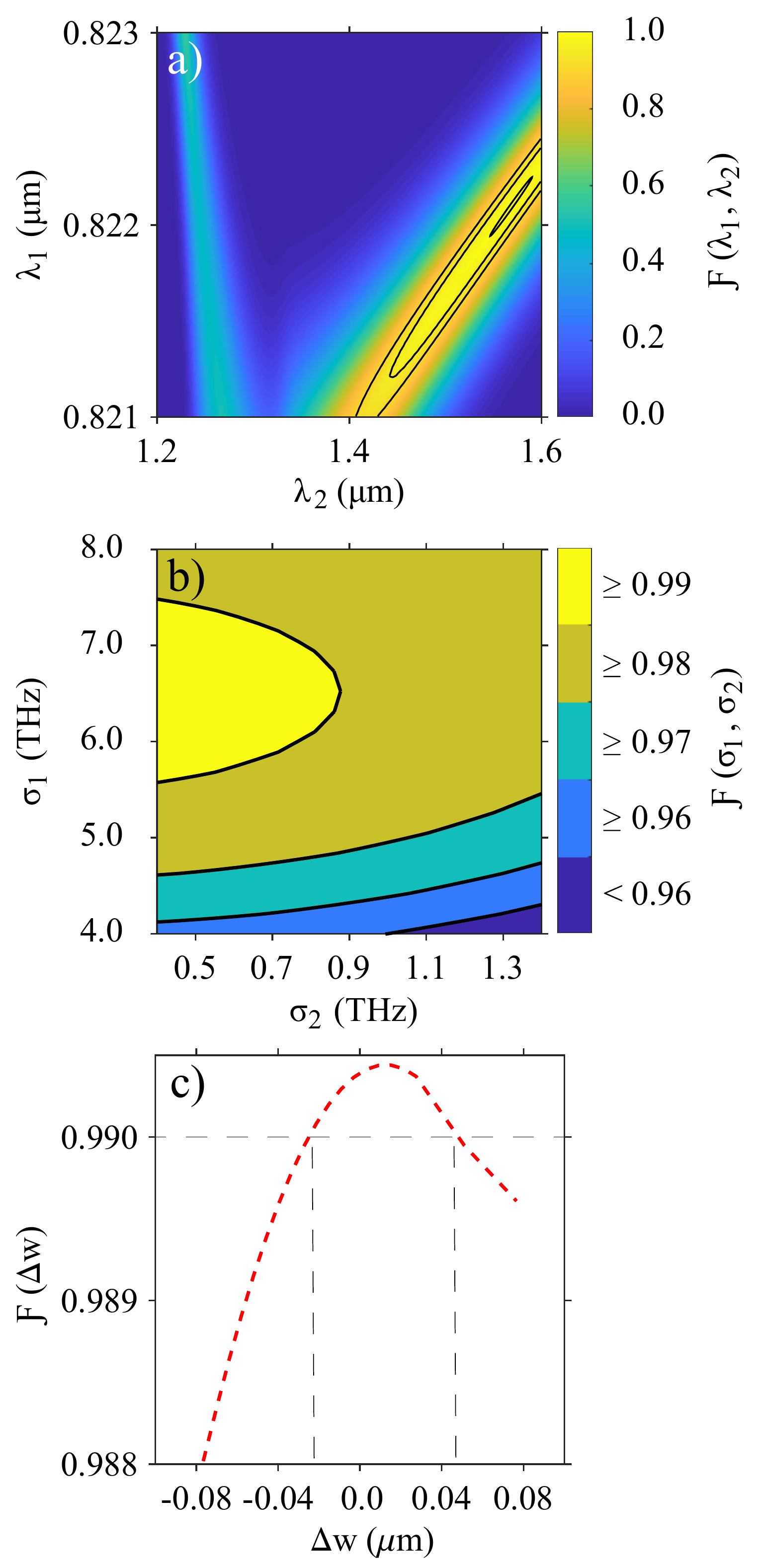}
  	\caption{\footnotesize{Color qubit preparation fidelity as a function of a) the pump wavelengths, b) the pump bandwidths, and c) variations in the SFWM and DFG waveguide widths, while the remaining parameters of the SFWM and DFG sources are fixed to the values in Table \ref{tab:parameter}.}}
  	\label{fig:FidLamSig}
  	\end{center}
  \end{figure}

\section*{DISCUSSION}

As described in the previous section, our designed integrated photonic circuit allows the preparation of color qubit with high fidelity. A continuum of linear superpositions can be prepared by changing the coupling parameter $\theta_j$, which can be achieved by externally controlling the average power of the two pump fields. From Equation (\ref{eqn:thetanPulCont_2}), and under our proposed experimental conditions, we estimate that a complete translation of the single-photon state from $\omega_s$ to $\omega_r$ is achieved for a minimal product of the pump powers of $8.3$ mW$^2$. A reasonable value to avoid distortion of signal spectra that can degrade our quantum gate's fidelity, related to other nonlinear effects in the waveguides \cite{Christ2013, Boyraz2004}.

Even though in the previously discussed example, we assumed an input state comprising components in just one dimension of the computational basis, in general, the input state to the DFG source may have the form $\ket{\Psi}_{in} =  \sum_i \left( x_i \hat{A}^\dagger_i + y_i \hat{B}_i^\dagger   \right) \vac$, for which the complete output state is calculated as $\ket{\Psi}_{out}=\textbf{R}_{\hat{\bold{n}}(\nu)}\ket{\Psi}_{in}$. Notice that the general input state can be prepared in an equivalent DFG source, as shown in Figure \ref{fig:Scheme}, suggesting an additional but feasible stage in the same integrated photonic circuit. Likewise, as mentioned in the Results section, our proposed qubit logic is constrained to rotations around axes in the xy plane of the Bloch sphere, which can also be implemented in our designed photonic circuits by modulating the phase difference between the two pump pulses through an active integrated optical component \cite{Sorianello2018}. However, note that entire Bloch sphere rotations could be realized by scaling our proposed device to allow consecutive rotations around the x and y axes, something entirely achievable by integrated photonic means, so its design is currently being explored.
  
Our proposed design, characterized by parameters given in Table \ref{tab:parameter},  represents a feasible device to manufacture and characterize using the working group infrastructure. Still, in practical implementations, deviation from predicted conditions may occur, leading to undesired results. We estimate the quantum gate fidelity value as a function of pump wavelengths tunability and bandwidths and variations in the SFWM and DFG waveguide widths to validate our design's robustness. Results are shown in Figure \ref{fig:FidLamSig}. Surprisingly, from panel a) we can see a continuum of pump wavelength pairs for which fidelity reaches values close to unity. The three highlighted contours from external to internal correspond to fidelities higher than 0.9, 0.95 and 0.99, respectively. Note that for a wide $\lambda_2$ range (from $1.4$ to $1.6$ $\mu$m), the fidelity remains high enough for an appropriate $\lambda_1$ value tunable in a tight, narrow range. As noticed, the function $\mathcal{F}(\lambda_1,\lambda_2)$ exhibits two prominent stripes that tend to converge to a particular $\lambda_2$ value. In this convergency point, the DFG phasematching nontrivial solution and the two trivial phasematching solutions merge. For a fixed $\lambda_1$, the fidelity can reach higher values for longer $\lambda_2$'s (right stripe) than for $\lambda_2$'s shifted to the visible (left stripe). As explained in the previous section, the fidelity depends on the spectral overlap between the eigenfunction $\phi_1(\omega_s)$ and $\varphi_1(\omega_s)$ as dictated by $\mathcal{O}_j$ (see Equation \ref{eq:overlap}). Consequently, lower or negligible fidelity values (Figure \ref{fig:FidLamSig}a)) correspond to $\lambda_1$ and $\lambda_2$ pairs, leading to a diminished or null spectral overlapping. Note that trivial solutions represent the two cases i) $\omega_1=\omega_2$ and then $\omega_s =\omega_r$, and ii) $\omega_s=\omega_2$, and then $\omega_r=\omega_1$. 

On the other hand, we evaluate the fidelity as a function of the bandwidths of the two pumps, $\mathcal{F}(\sigma_1,\sigma_2)$, while keeping the remaining parameters as described in Table \ref{tab:parameter}. Results are displayed in Figure \ref{fig:FidLamSig}b), where it is seen that fidelity values close to unity are obtained for narrower $\sigma_2$ values in the considered bandwidth range and $\sigma_1$ ranging approximately between $5.5$ and $7.5$ THz. Variations in the pump bandwidths affect the fidelity because the SFWM and DFG spectra change, leading to optimal or non-optimal spectral overlaps between the pair of eigenfunctions $\phi_1(\omega_s)$ and $\varphi_1(\omega_s)$, see Figure \ref{fig:spectra} and Equation (\ref{eq:overlap}).

Figure \ref{fig:FidLamSig}c) shows the fidelity as a function of variations in the SFWM and DFG waveguide widths, rewritten as $w_{dfg} \rightarrow$ $w_{dfg} + \Delta w$ and $w_{sfwm} \rightarrow$ $w_{sfwm} + \Delta w$, with $\Delta w$ the deviation from the values reported in Table \ref{tab:parameter}. We assume a simultaneous increase or decrease of the waveguide widths that can be originated, for instance, from a resolution variation in the system used to write the waveguides. The maximum width deviation considered in the simulation is $\Delta w = \pm 0.076 \, \mu$m. To calculate the fidelity, we fixed the other source parameters to the values reported in Table \ref{tab:parameter}, except the operation wavelengths $\lambda_1$, $\lambda_s$, $\lambda_i$, $\lambda_r$, which must be adjusted for simultaneous perfect phasematching of the SFWM and DFG processes as waveguide dispersion properties modify with changes in the waveguide widths. It can be seen in the figure that fidelity remains close to unity for the explored $\Delta w$ range. Note there is an interval with a $0.07\,\mu$m width for which the fidelity is greater than $0.99$, defined by the vertical dashed lines in Figure \ref{fig:FidLamSig}c). The described analysis shows that our proposed design is robust against fabrication variations. Although the fidelity value varies if the waveguide widths result broader or thinner, the change is not so dramatic in an interval of more than $100$ nm, while the operation wavelengths of the device remain experimentally accessible.

Results depicted in Figure \ref{fig:FidLamSig} reveal that once the proposed circuit has been implemented, there is flexibility in controlling its operation through external parameters related to the laser sources. They also indicate that if the geometrical parameters deviate slightly from the ideal values during the fabrication process, suitable quantum gate fidelities can still be obtained by choosing the proper phasematched wavelengths and pump bandwidth; a consequence of the SFWM and DFG spectral properties dependence on waveguide dispersion and pump characteristics.
  
 \begin{figure*}[t!]
\begin{center}
\includegraphics[width=0.97\linewidth]{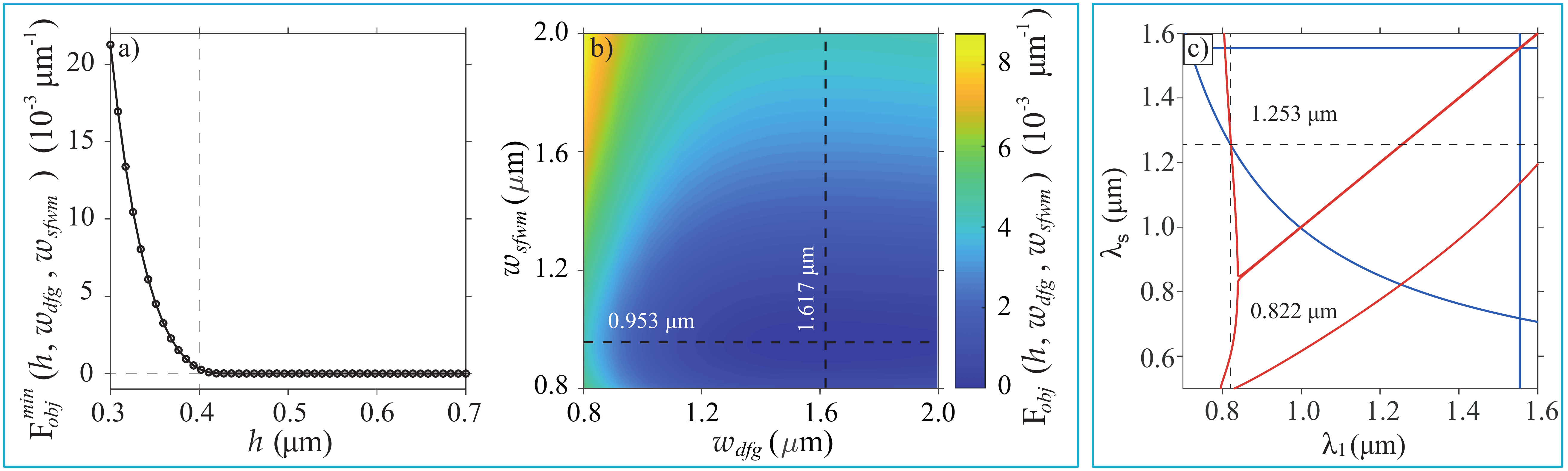}
\caption{\footnotesize{ Method for simultaneous SFWM and DFG phasematching. a) The minimum value of the objective function in terms of the core height $h$, when the widths of the two waveguides are allowed to vary. b) Value of the objective function for $h = 0.7\,\mu $m in terms of the two waveguide widths. c) Phasematching contours for SFWM (red) and DFG (blue), obtained with parameters $ h = 0.7\, \mu $m, $w_ {sfwm} = 0.953\,\mu $m and $ w_ { dfg} = 1.617 \,\mu $m.}}
\label{fig:FnObj}
\end{center}
\end{figure*}

In conclusion, we have designed a feasible integrated photonic circuit capable of color qubit preparation and rotations, based on the third-order nonlinear processes of spontaneous four-wave mixing and difference frequency generation. Our proposed device settles the basis for a photonic single-qubit logic relying on the temporal modes of pulsed single-photon states. It has been proved that by spectral engineering of the two nonlinear processes, assumed to be implemented in silicon nitride rectangular waveguides, qubit preparation fidelities close to unity are attainable. We consider our work a viable route for advancing quantum computing development by photonic means.

\section*{METHODS}

For designing our integrated photonic circuit proposed to demonstrate color qubit preparation and rotations by DFG, we developed a parametric numerical exploration that enabled us to identify the set of physical parameters that lead to a quantum gate fidelity as close to unity as possible. For this, we propose an interaction in which the fundamental temporal modes of the input state and DFG quantum gate spectra almost perfectly overlap. This strategy requires satisfying the following conditions: i) the JSA must be factorable, such that the heralded single-photon state contains only the fundamental mode, and ii) although the mapping function can comprise several temporal modes, it must be such that only the fundamental one overlaps with that of the input state. Ideally, the overlap must be perfect for getting a unity fidelity, so the first step in our quest is to guarantee the simultaneous fulfillment of the phasematching conditions for the two processes (SFWM and DFG) at the same central frequencies $\omega_1$ and $\omega_s$, which depends on the dispersion properties of the nonlinear media. 

  \begin{figure}[t!]
	\begin{center}
		\includegraphics[width=0.98\linewidth]{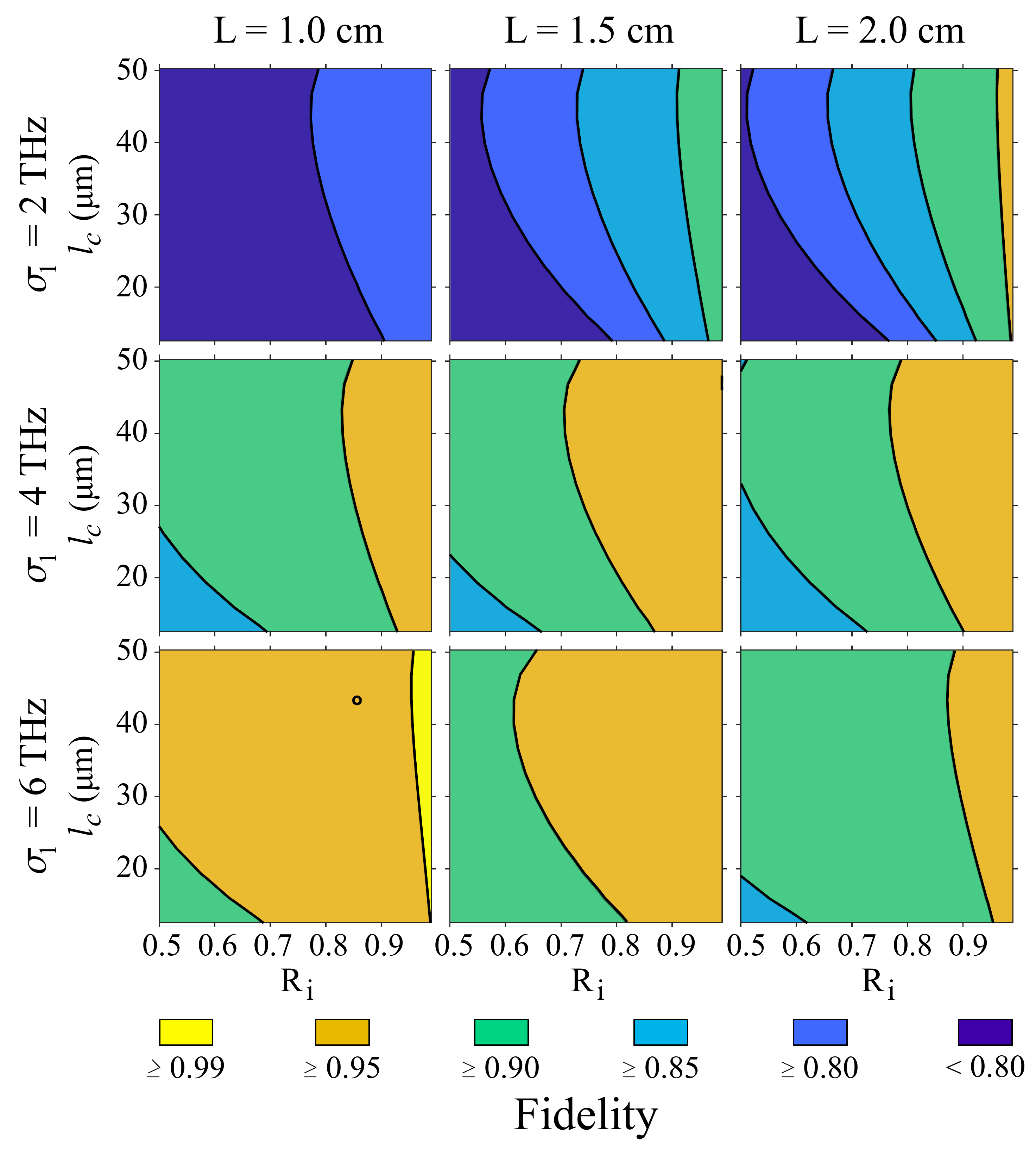}
		\caption{\footnotesize{Methodology for increasing the qubit preparation fidelity. Single-qubit preparation fidelity in terms of the length of the DFG waveguide ($L$), the bandwidth of pump 1 ($\sigma_1$ ), and the length ($l_c$ ) and reflectivity ($R_i$) of the SFWM cavity, for $\sigma_2 = 0.5$ THz and $\sigma_f=4$ THz. The black circle marker indicates the combination of parameters we choose for our example.}}
		\label{fig:FidMatrix}
	\end{center}
\end{figure}  

Note that the material optical properties used in our numerical study correspond to those materials synthesized and characterized by the working group \cite{Aguayo2020a}, which become essential to make a real connection among simulations, experimental data, and feasible proposals. On the other hand, the waveguide dispersion properties were computed employing the opensource optical eigenmode solver WGMODES \cite{Fallahkhair2008}, by using as free parameters the width ($w$) and the height ($h$) of the waveguide cores, while the height of the silicon dioxide layer was set to $1\, \mu$m.

\begin{figure}[t!]
	\begin{center}	\includegraphics[width=0.94\linewidth]{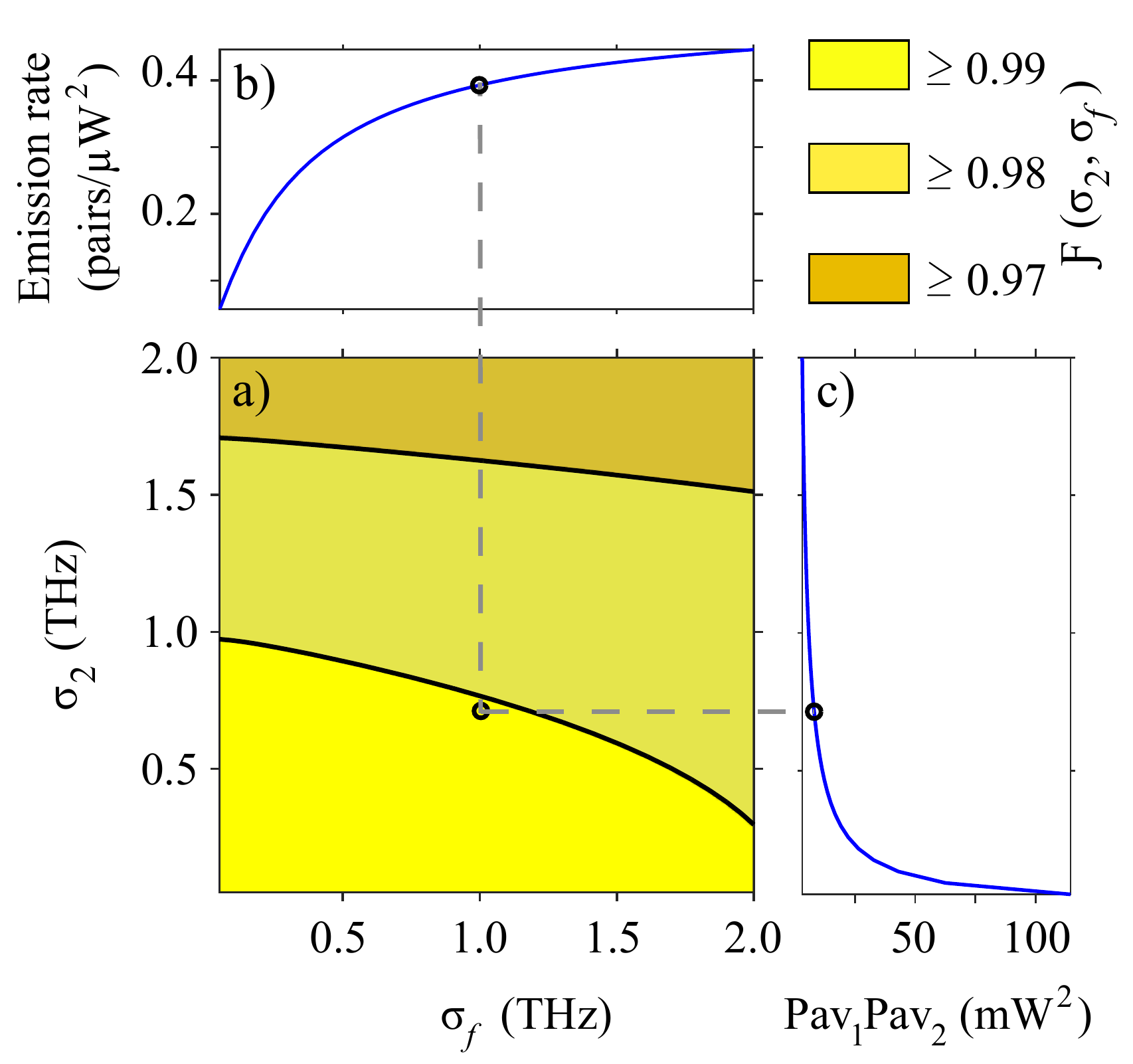}
		\caption{\footnotesize{Methodology for increasing the qubit preparation fidelity. a) Fidelity as a function of $\sigma_2$ and $\sigma_f$, while fixing the remaining parameters at the values corresponding to the black circle marker in Figure \ref{fig:FidMatrix}. b) The photon-pair emission rate per pump-power squared in terms of $\sigma_f$. c) Minimum product of the average powers of the two pump fields required for a complete frequency conversion in terms of $\sigma_2$. The black circle markers correspond to the choice for our design.}}
	\label{fig:FilterVar}
	\end{center}
\end{figure}

To accomplish the conditions mentioned above, we proceed with the following methodology:

\begin{enumerate}[label=(\roman*)]
\item To define the pump and signal wavelengths for the SFWM and DFG processes. In our case, taking into account our concern in an experimental demonstration of the proposed integrated circuit, we choose the central operation wavelengths of the device based on our availability of lasers sources and single-photon detectors for characterization. 
\item To identify the waveguide geometrical parameters for the SFWM and DFG sources. For doing so, we define an objective function to be minimized, given by 
\begin{align}
	F_{obj}(h,w_{sfwm},w_{dfg}) &= \Delta k_{sfwm}^2(h,w_{sfwm}) \nonumber \\
	&+\Delta k_{dfg}^2(h,w_{dfg}),
	\label{eqn:funObj}
\end{align}

\noindent where $\Delta k_{sfwm}(h,w_{sfwm})$ $[\Delta k_{dfg}(h,w_{dfg})]  $ represents the phase mismatching for SFWM [DFG]. Perfect phasematching requires the conditions $\Delta k_ {sfwm} (h, w_ {sfwm}) = 0$ and $\Delta k_ {dfg} (h, w_ {dfg}) = 0$ to be fulfilled, which reduces our problem to find the values of $h$, $w_ {sfwm}$ and $w_ {dfg}$ that minimize the objective function at the wavelengths of interest. We consider that fields in each waveguide propagate in the fundamental electrical transverse mode. The exploration ranges for each dimension were limited to realistic and feasible manufacturing values under our experimental conditions. Figure \ref{fig:FnObj}a) shows the minimum value of the objective function in terms of the core height $h$, while the widths of the two waveguides are allowed to vary. It is interesting to see that for a certain value of the core height, $h\approx 0.4\,\mu$m, the function $F_ {obj} (h, w_ {sfwm}, w_ { dfg})$ reaches a plateau near the desired value $F_ {obj} (h, w_ {sfwm}, w_ {dfg}) = 0$. The observed behavior suggests that for each $h$ on that plateau, there is a combination of $w_ {sfwm}$ and $w_ {dfg}$ values, for which phasematching is simultaneously achieved for the two processes. Figure \ref{fig:FnObj}b) shows the value of the objective function for $h = 0.7\,\mu $m in terms of the waveguide widths. It is worth noting that the minimum value shifts to lower $w_ {dfg}$ as the height decreases, while $w_ {sfwm} $ remains almost constant. Figure \ref{fig:FnObj}c) shows the phasematching diagrams for the two processes, obtained with parameters $ h = 0.7\, \mu $m, $w_ {sfwm} = 0.953\,\mu $m and $ w_ { dfg} = 1.617 \,\mu $m.  In the case of DFG the diagram is generated for $\lambda_ {2} = 1.554\,\mu $m. This figure shows that phasematching is fulfilled simultaneously for the two processes at $\lambda_1=0.822 \,\mu$m and $\lambda_s=1.253 \,\mu$m.

\item To boost the quantum gate fidelity through variations of user-accessible parameters. By keeping fixed the bandwidths of pump 2 ($\sigma_2 = 0.5$ THz) and the spectral filter in the heralding arm of the SFWM source ($\sigma_f=4$ THz), we evaluate the fidelity as a function of the remaining device parameters in order to identify conditions for boosting the fidelity value. The free variables were the length of the DFG waveguide ($L$), the bandwidth of pump 1 ($\sigma_1$), and the length ($l_c$) and reflectivity ($R_i$) of the SFWM cavity, according to the model in reference \cite{Garay2011}. The parameters were varied in ranges of values accessible experimentally. Results are displayed in Figure 5 as a $3\times3$ matrix, where columns correspond to three different values of $L$, while rows correspond to three different values of $\sigma_1$. Each matrix element shows the fidelity value as a function of $l_c$ and $R_i$. Results are presented as filled contours, with six levels indicating different fidelities ranges; see the colorbar underneath the matrix. As the figure shows, in our conditions, the fidelity increases for shorter $L$, wider $\sigma_1$, longer $l_c$, and high-quality cavities (as dictated by the $R_i$ value); those parameter combinations favor the spectral overlap between the fundamental temporal modes of the input state and the mapping function characterizing the quantum gate medium. We choose the parameters corresponding to the black circle marker on the bottom-left matrix element for the example discussed here and presented as a potential proposal to be implemented, i.e., $L=1$ cm, $\sigma_1=6$ THz, $l_c=43\,\mu$m and $R_i =0.86$, which lead to a fidelity $\mathcal{F}=0.983$. 

As the last step in defining our design, we vary the bandwidths $\sigma_2$ and $\sigma_f$ while fixing the remaining parameters at the values corresponding to the black circle marker in Figure \ref{fig:FidMatrix}. Results are shown in Figure \ref{fig:FilterVar}. In addition to fidelity (panel a)), we present results of the SFWM photon-pair emission rate (panel b)) and the minimum product of the average powers of the two pump fields (panel c)) required for a complete evolution from the mode $\hat{A}^\dagger$ to the mode $\hat{B}^\dagger$, see Equations (\ref{eqn:AOper_1}) and (\ref{eqn:BOper_1}), since they are affected by $\sigma_2$ and $\sigma_f$. A high photon-pair emission rate is beneficial for getting reasonable heralding efficiencies of the SFWM based single-photon source, and a low minimum value of the product $P_{av1}P_{av2}$ is required for avoiding undesirable nonlinear effects in the DFG waveguide. Note that in the exploration ranges, the obtained fidelity values exceed 0.97 and can achieve values close to unity for smaller $\sigma_2$ and $\sigma_f$ at the expense of a reduced photon-pair emission rate and higher values of the product of the pump powers. For our proposed design, we set  $\sigma_2=0.7$ THz and $\sigma_f = 1$ THz, which lead to a fidelity $\mathcal{F}=0.99$, a photon-pair emission rate per pump-power squared of $\approx0.4$ pairs/$\mu$W$^2$ and a minimum product of the pump powers of $8.3$ mW$^2$. In Figure \ref{fig:FilterVar}, the black circle marker indicates this particular result. 
\end{enumerate}

The above-described methodology allowed us to find a feasible design with high potential to be developed with currently available technology for photonic integrated circuits. The full description of our proposed circuit for qubit preparation and rotations is summarized in Table 1.

\section*{DATA AVAILABILITY}
Data underlying the results may be obtained from the authors upon reasonable request.

\bibliographystyle{unsrt}

\begin{thebibliography}{80}


\bibitem{Debnath2016}
Debnath, S. et al. Demonstration of a small programmable quantum computer with atomic qubits. \textit{Nature} \textbf{536,} 63--66 (2016).    

\bibitem{Loss1998}
Loss, D.  $\&$  DiVincenzo, D. P. Quantum computation with quantum dots. \textit{Phys. Rev. A } \textbf{57,} 120 (1998).     

\bibitem{Uppu2020}
Uppu, R. et al. Scalable integrated single-photon source. \textit{Sci. Adv.} \textbf{6,} eabc8268 (2020).       

\bibitem{Song2017}
Song, C. et al. 10-Qubit Entanglement and Parallel Logic Operations with a Superconducting Circuit. \textit{Phys. Rev. Lett.} \textbf{119,} 180511 (2017).       

\bibitem{Ladd2010}
Ladd, T. D. et al. Quantum computers. \textit{Nature} \textbf{464,} 45--53 (2010). 

\bibitem{Mosley2008}
Mosley, P. J. et al. Heralded Generation of Ultrafast Single Photons in Pure Quantum States. \textit{Phys. Rev. Lett.} \textbf{100,} 133601 (2008).     

\bibitem{Soller2011}
S\"oller, C., Cohen, O., Smith, B. J., Walmsley, I. A. $\&$ Silberhorn, C. High-performance single-photon generation with commercial-grade optical fiber. \textit{Phys. Rev. A} \textbf{83,} 031806(R) (2011).     

\bibitem{DiVicenzo2000}
DiVicenzo, D. P. The Physical Implementation of Quantum Computation. \textit{Fortschritte der Phys.} \textbf{48,} 9--11 (2000). 

\bibitem{Qiang2018}
Qiang, X. et al. Large-scale silicon quantum photonics implementing arbitrary two-qubit processing. \textit{Nat. Photonics} \textbf{12,} 534--539 (2018). 

\bibitem{Zhukovsky2012}
Zhulovsky, S. V. et al. Generation of maximally-polarization-entangled photons on a chip. \textit{Phys. Rev. A} \textbf{85,} 013838 (2012). 

\bibitem{Signorini2020}
Signorini, S. $\&$ Pavesi, L. On-chip heralded single photon sources. \textit{AVS Quantum Sci.} \textbf{2,} 041701 (2020). 

\bibitem{Wang2019}
Wang, Y. et al. Efficient quantum memory for single-photon polarization qubits. \textit{Nat. Photonics} \textbf{13,} 346--351 (2019). 

\bibitem{Crespi2011}
Crespi, A. et al. Integrated photonic quantum gates for polarization qubits. \textit{Nat. Commun} \textbf{2}, 566 (2011). 

\bibitem{Pewkhom2020} Pewkhom, P., Suwanna S. $\&$ Kalasuwan, P. Alternative scheme of universal optical programmable multi-qubit gates for polarization qubits. \textit{Quantum Inf. Process} \textbf{19} 209 (2020). 

\bibitem{Laplane2016} Laplane, C. et al. Multiplexed on-demand storage of polarization qubits in a crystal. \textit{New J. Phys.} \textbf{18}, 013006 (2016).


\bibitem{Nicolas2014}
Nicolas, A. et al. A quantum memory for orbital angular momentum photonic qubits. \textit{Nat. Photonics} \textbf{8,} 234--238 (2014). 

\bibitem{Raymer2020}
Raymer, M. G. $\&$ Walmsley, I. A. Temporal modes in quantum optics: then
and now. \textit{Phys. Scr.} \textbf{95,} 064002 (2020). 

\bibitem{Brecht2015}
Brecht, B., Reddy, D. V., Silberhorn, C. $\&$ Raymer, M. G. Photon Temporal Modes: A Complete Framework for Quantum Information Science. \textit{Phys. Rev. X} \textbf{5,} 041017 (2015). 

\bibitem{Ashby2020}
Ashby, J. et al. Temporal mode transformations by sequential time and frequency phase modulation for applications in quantum information science. \textit{Opt. Express} \textbf{28,} 38376--38389 (2020). 

\bibitem{Eckstein2011}
Eckstein, A., Brecht, B. $\&$ Silberhorn, C. A quantum pulse gate based on spectrally engineered sum frequency generation. \textit{Opt. Express} \textbf{19,} 13770--13778 (2011). 

\bibitem{Brecht2011}
Brecht, B., Eckstein, A., Christ, A., Suche, H. $\&$ Silberhorn, C. From quantum pulse gate to quantum pulse shaper-engineered frequency conversion in nonlinear optical waveguides. \textit{New J. Phys.} \textbf{13,} 065029 (2011). 

\bibitem{Allgaier2020}
Allgaier, M. et al. Pulse shaping using dispersion-engineered difference frequency generation. \textit{Phys. Rev. A} \textbf{101,} 043819 (2020). 


\bibitem{Knill2001}
Knill, E., Laflamme, R. $\&$ Milburn, G. J. A scheme for efficient quantum computation with linear optics. \textit{Nature} \textbf{409,} 46--52 (2001). 

\bibitem{Maring2018} 
N. Maring, 
\textit{Quantum Frequency Conversion for Hybrid Quantum Networks}, 
PhD Thesis dissertation, Universitat Polit\'ecnica de Catalunya, Barcelona, Espa\~na (2018).    

\bibitem{McKinstrie2005}
McKinstrie, C. J., Harvey,  J. D., Radic, S. $\&$ Raymer, M. G. Translation of quantum states by four-wave mixing in fibers. \textit{Opt. Express} \textbf{13}, 9131--9142 (2005). 

\bibitem{McGuinness2010}
McGuinness, H. J., Raymer, M. G., McKinstrie, C. J. $\&$ Radic, S. Quantum Frequency Translation of Single-Photon States in a Photonic Crystal Fiber. \textit{Phys. Rev. Lett.} \textbf{105,} 093604 (2010).    

\bibitem{Clemmen2016}
Clemmen, S., Farsi, A., Ramelow, S. $\&$ Gaeta, A. L. Ramsey interference with single photons. \textit{Phys. Rev. Lett.} \textbf{117,} 223601 (2016).

\bibitem{Dominguez2021}
Dominguez-Serna, F. $\&$ Garay-Palmett, K. Quantum state preparation and one qubit logic from third-order nonlinear interactions. \textit{J. Opt. Soc. Am. B} \textbf{38,} 2277--2283 (2021). 

\bibitem{Fewings2000}
Fewing, M. R. $\&$ Gaeta, A. L. Compensation of pulse distortions by phase conjugation via difference-frequency generation. \textit{J. Opt. Soc. Am. B} \textbf{17,} 1522--1525 (2000).    


\bibitem{Grice2001}
Grice, W. P., U Ren, A. B. $\&$ Walmsley, I. A.  Eliminating frequency and space-time correlations in multiphoton states. \textit{Phys. Rev. A} \textbf{64,} 063815 (2001). 

\bibitem{Garay2007}
Garay-Palmett, K. et al. Photon pair-state preparation with tailored spectral properties by spontaneous four-wave mixing in photonic-crystal fiber. \textit{Opt. Express} \textbf{15,} 14870--14886 (2007). 

\bibitem{Reddy2014}
Reddy, D. V., Raymer, M. G. $\&$ McKinstrie, C. J. Efficient sorting of quantum-optical wave packets by temporal-mode interferometry. \textit{Optics Lett.} \textbf{39,} 2924--2927 (2014). 

\bibitem{Reddy2015}
Reddy, D. V. $\&$ Raymer, M. G. $\&$ McKinstrie, C. J. Sorting photon wave packets using temporal-mode interferometry based on multiple-stage quantum frequency conversion. \textit{Phys. Rev. A} \textbf{91,} 012323 (2015). 

\bibitem{Quesada2016}
Quesada, N. $\&$ Sipe, J. E. High efficiency in mode-selective frequency conversion. \textit{Optics Lett.} \textbf{41,} 364--367 (2016). 

\bibitem{Reddy2018}
Reddy, D. V. $\&$ Raymer, M. G. High-selectivity quantum pulse gating of photonic temporal modes using all-optical Ramsey interferometry. \textit{Optica} \textbf{5,} 423--428 (2018). 

\bibitem{Aguayo2020a}
Aguayo-Alvarado, A. L., Acevedo-Carrera, A., Dominguez-Serna, F. A., De La Cruz, W. $\&$ Garay-Palmett, K. A Proposal for Nonlinear-Optics Based Quantum Gates in Integrated Photonic Circuits. \textit{Frontiers in Optics} \text{FM4A. 8} (2020).   

\bibitem{Fu2014}
Fu, Y., Ye, T., Tang, W. $\&$ Chu, T. Efficient adiabatic silicon-on-insulator waveguide taper. \textit{Photon. Res.} \textbf{2,} A41--A44  (2014).

\bibitem{Garay2011}
Garay-Palmett, K., Jeronimo-Moreno, Y. $\&$ U Ren, A. B. Theory of cavity-enhanced spontaneous four wave mixing. \textit{Laser Phys.} \textbf{23,} 015201 (2013). 

\bibitem{Nielsen2010}
Nielsen, M. A. $\&$  Chuang, I. L. Quantum Computation and Quantum Information: 10th Anniversary Edition (Cambridge University Press, Cambridge, 2013). 

\bibitem{Dominguez2020}
Dominguez-Serna, F. A., Rojas, F. $\&$ Garay-Palmett, K. Quantum teleportation with hybrid entangled resources prepared from heralded quantum states. \textit{J. Opt. Soc. Am. B}  \textbf{37,} 695--701 (2020).        

\bibitem{Law2000}
Law, C. K., Walmsley, I. A. $\&$ Eberly, J. H. Continuous Frequency Entanglement: Effective Finite Hilbert Space and Entropy Control. \textit{Phys. Rev. Lett.}  \textbf{84,} 5304 (2000).      

\bibitem{Xu2021}
Xu, S. T. et al. Intensity-tunable terahertz bandpass filters based on liquid crystal integrated metamaterials. \textit{Appl. Opt.}  \textbf{60,} 9530--9534  (2021).

\bibitem{Zhang2022}
Zhang, B. et al. Bandwidth Tunable Optical Bandpass Filter Based on Parity-Time Symmetry. \textit{Micromachines} \textbf{13,} (1) 89  (2022).

\bibitem{Nasr2021}
Nasr, N., Younes, A. $\&$ Elsayed, A. Efficient representations of digital images on quantum computers. \textit{Multimed Tools Appl}  \textbf{80,} 34019--34034 (2021).      

\bibitem{Christ2013}
Christ, A., Brecht, B., Mauerer, W. $\&$ Silberhorn, C. Theory of quantum frequency conversion and type-II parametric down-conversion in the high-gain regime \textit{New J. Phys.}  \textbf{15,} 053038 (2013).      

\bibitem{Boyraz2004}
Boyraz, O., Indukuri, T. $\&$ Jalali, B. Self-phase-modulation induced spectral broadening in silicon waveguides. \textit{Opt. Express}  \textbf{12,} 829--834 (2004).      

\bibitem{Sorianello2018}
Sorianello, V. et al. Graphene–silicon phase modulators with gigahertz bandwidth. \textit{Nat. Photonics}  \textbf{12,} 40--44 (2018).      

\bibitem{Fallahkhair2008}
Fallahkhair, A. B., Li, K. S. $\&$ Murphy, T. E. Vector Finite Difference Modesolver for Anisotropic Dielectric Waveguides. \textit{J. Lightwave Technol.} \textbf{26,} 1423-1431 (2008). 

\end{thebibliography}

\section*{ACKNOWLEDGEMENTS}
K. G-P. acknowledges funding support by Consejo Nacional de Ciencia y Tecnología (CONACYT) (grants FORDECYT-PRONACES/194758, FORDECYT-PRONACES 298971 and CATEDRAS-CONACYT 709/2018). A.L.A-A. is supported by a CONACYT Fellowship. F.D-S. acknowledges support by CONACYT(grant CATEDRAS-CONACYT 709/2018). W.D.L.C. acknowledges funding support by CONACYT (grant FORDECYT-PRONACES/194758), DGAPA-UNAM (grant IT101021) and the Laboratorio Nacional de Nanofabricaci\'on of CNyN-UNAM and CONACYT. We also thank WGMODES developers for the open availability of their Optical Eigenmode Solver.

\section*{AUTHOR CONTRIBUTIONS}
A.L.A-A. carried out the theoretical and numerical studies and the synthesis and optical characterization of materials considered for waveguide simulations. F.D-S. conducted the theoretical study and contributed to the analysis of the results, W.D.L.C. conducted the optical characterization of materials for waveguides and analyzed the corresponding results, K. G-P. conducted the integrated photonic circuit design and contributed to the analysis of the results. All authors reviewed the manuscript.

\section*{COMPETING INTERESTS}
The authors declare no competing interests.

\end{document}